\documentclass{elsart}
\usepackage{amsmath,amssymb}
\newcommand{\tr}{\mathop{\mathrm tr}}

\newcommand{\ii}{\mathrm{i}}
\newcommand{\pd}{\partial}
\newcommand{\I}{\mathbb{I}}
\newcommand{\ket}[1]{|#1\rangle}
\newcommand{\bra}[1]{\langle #1|}
\newcommand{\Sf}{S_{\mathrm{f}}}
\newcommand{\Sn}{S_{\mathrm{naive}}}

\begin{document}

\runauthor{eu}
\begin{frontmatter}
\title{Matrix Models: Fermion Doubling vs. Anomaly\thanksref{X}}

\thanks[X]{Work supported by RFBR
grant \# 99-01-00190, INTAS grant \# 950681, and Scientific School
support grant 96-15-0628}

\author[Chisinau,Dubna]{Corneliu Sochichiu\thanksref{e-mail1}}
\address[Chisinau]{Institutul de Fizic\u a Aplicat\u a
A\c S, str. Academiei, nr. 5, Chi\c sin\u au, MD2028 MOLDOVA}
\address[Dubna]{ Bogoliubov Laboratory of
Theoretical Physics, Joint Institute for Nuclear Research, 141980
Dubna, Moscow Reg., RUSSIA\thanksref{Present}}
\thanks[e-mail1]{E-mail: \texttt{sochichi@thsun1.jinr.ru}}
\thanks[Present]{Present address.}
\begin{abstract}
We present some arguments showing spectrum doubling of matrix
models in the limit $N\rightarrow\infty$ which is connected with
fermionic determinant behaviour. The problems are similar to ones
encountered in the lattice gauge theories with chiral fermions.
One may discuss the ``physical meaning'' of the doubling states or
ways to eliminate them. We briefly consider both situations.
\end{abstract}
\end{frontmatter}
\typeout{SET RUN AUTHOR to \@runauthor}

\section{Introduction}
A tool for nonperturbative study of superstring theory (M-theory)
proved to be matrix models, in special BFSS (M(atrix) theory) and
IKKT (IIB matrix theory) ones, \cite{ikkt,Banks:1997vh}. These
models are formulated in terms of $N\times N$ Hermitian matrices,
where $N\rightarrow\infty$.

For a finite (but large) $N$ the last model, \cite{ikkt} plays the
r\^{o}le of regularised ``second quantised'' IIB superstring in
Schild formulation \cite{Schild:1977vq}, while the former one
\cite{Banks:1997vh} is a regularisation of $D=11$ quantum
super-membrane.

Even for finite $N$ these models have problems with definiteness
due to divergence of the partition function arising from
integration along flat directions or vacua. Due to this also the
limit $N\rightarrow\infty$ is bad defined. However, if not to try
to consider the matrix model as a whole, but to consider a
``perturbative sector'' connected with fluctuations around a
particular vacuum solution one may have a well defined system.

The models obtained this way depend on the chosen vacuum solution,
in particular, various compactifications of IIB matrix model on
noncommutative tori were shown to yield super Yang--Mills (sYM)
models on these tori, as well as BFSS model, \cite{Connes:1998cr}.

One can show that in fact Yang--Mills type models describe all
regular fluctuations around a vacuum solution of IKKT matrix
model, \cite{Sochichiu:2000eu}. Under regular fluctuations we
understand those which have bounded value of momentum and position
operators. In practice e.g. numerical computations this
restriction can be implemented by additional regularisation which
makes the undesirable modes decouple.

Once there is a close correspondence between matrix models and the
sYM models there appears the following problem. Ten dimensional
sYM model is known to be anomalous and the anomaly seems to exist
also in the noncommutative case,
\cite{Gracia:2000eu,Bonora:2000eu}. From the other hand there is
no visible source for anomaly in the matrix model. For any finite
$N$, the matrix model is finite manifestly gauge and Lorentz
invariant as well as supersymmetric. Naively, these properties
must hold also in the limit $N\rightarrow\infty$, contradicting
the anomaly of $D=10$ sYM.

The explanation of this discrepancy may reside just in the
contribution of the singular configurations. In the actual paper
we show that the matrix model fails to reproduce the anomaly in
the limit $N\rightarrow\infty$ due to the spectrum doubling in the
fermionic sector of the matrix model.

In fact the actual situation is not new. A class of models is known to suffer from
the spectrum doubling, \cite{Creutz:1996wq,Kits}. In particular, for lattice gauge
theory, it is known that the the naive discretisation of chiral fermionic action
leads to doubling of the fermionic spectrum. The last results in appearance for
each mode another mode(s) carrying the opposite chirality which compensate the
parity odd contribution of fermions, \cite{Wilson:1974sk}. In fact, there is a
no-go theorem due to Nielsen and Ninomiya \cite{Nielsen:1981xu,Nielsen:1981rz},
which states that the doubling cannot be avoided unless the gauge, Lorentz or
other relevant symmetry is destroyed in the continuum limit. While the gauge
symmetry plays the crucial r\^{o}le in the consistency of the model, breaking of
the Lorentz symmetry is not dangerous for the model and maybe even desirable as a
mechanism of spontaneous breaking of the ten dimensional Lorentz group to a lower
dimensional one,~\cite{NV}.

The plan of the paper is as follows. First we introduce the
description of the fermionic sector of IIB matrix model for finite
$N$, after that we explicitly find the doubling states for free
fermionic fluctuations (quadratic approximation), and discuss the
issue for the interacting case as well as possibility to eliminate
the doubling.

\section{Finite $N$ Matrix Model}

The IIB matrix model is described by the action,
\begin{equation}\label{action}
  S=-\frac{1}{g^2}\tr \left(\frac{1}{4}[A_\mu,A_\nu]^2+
  \frac{1}{2}\bar{\psi}\Gamma^\mu[A_\mu,\psi]\psi\right),
\end{equation}
where $A_\mu$ and $\psi$ are ten dimensional vector and
Majorana--Weyl spinor Hermitian $N\times N$ matrices.

An important class of configurations which tend to satisfy
equations of motion in the limit $N\rightarrow\infty$, are BPS
ones given by $A_\mu=p_\mu$, where Hermitian matrices
$p_\mu=p_\mu^{\dag}$ satisfy,
\begin{equation}\label{eq:B}
  [p_\mu,p_\nu]=\ii B_{\mu\nu},
\end{equation}
here $B_{\mu\nu}$ is proportional to the unity matrix
$B_{\mu\nu}\equiv B_{\mu\nu}\cdot \I$.

Although, such a set of matrices does not exist for finite $N$, it
can be approximated by a sequence of matrices converging to
(\ref{eq:B}) in the sense of operator norm on the Hilbert space of
smooth finite functions (vectors), \cite{Sochichiu:2000eu}.

Before searching such an approximation, consider a Lorentz
transformation which brings matrix $B_{\mu\nu}$ to the canonical
form, having $2\times 2$ antisymmetric diagonal blocks with $\pm
\hbar_i$ entries. In this case the set of matrices $p_\mu$ is
split in pairs $(p_i,q^i)$, $i=1,\dots,D/2$, where $p_i$ and $q^i$
are canonically conjugate,
\begin{align}\label{pp}
  &[p_i,p_j]=[q^i,q^j]=0, \\ \label{pq}
  &[p_i,q^j]=-\ii \hbar_i\delta_i^j.
\end{align}

Eqs. (\ref{pp},\ref{pq}) give the $D/2$ dimensional Heisenberg
algebra. It is known that the Heisenberg algebra can be
represented e.g. in terms of square integrable functions defined
on on the spectrum of operators $q^i$.

Finite $N$ analog of the Heisenberg algebra (\ref{pq}) is given by
the position and the Hermitian (symmetric) lattice derivative
operator on a compact rectangular periodic lattice $\Gamma$. Let
$\ket{n}$ be eigenvector of $q^i$ with eigenvalues $L_i\sin
({2\pi}/{N_i})$, $\prod_i N_i=N$,
\begin{equation}\label{qi}
  q^i\ket{n}=L_i\sin(2\pi n^i/N_i)\ket{n},
\end{equation}
then operators $p_i$ act on this basis as follows,
\begin{equation}\label{pn}
  p_i\ket{n}=\frac{\ii\hbar_i}{2a_i}\left(\ket{n+e_i}-
  \ket{n-e_i}\right),
\end{equation}
where $e_i$ is the unity lattice vector along $i$-th link, and
$a_i=\frac{2\pi L_i}{N_i}$. The choice of $q$ which we use differs
from one usually used in lattice model by redefinition of
$q\rightarrow L\sin 2\pi q/L$. It is as good as the former one,
but beyond this it treats $p$ and $q$ in a symmetric manner. It is
not difficult to see that in the basis of $p$ eigenvectors eqs.
(\ref{qi}) and (\ref{pn}) keep the same form with $p$ and $q$
interchanged. It also introduces explicitly the periodic boundary
conditions, i.e. the identification $\ket{n+N_ie_i}\sim \ket{n}$.
Due to the symmetry between $p$ and $q$ the continuum limit is
achieved after the ``UV cutoff'' removing: $a_i\rightarrow 0$,
combined with the ``IR cutoff'' removing or
``decompactification'': $L_i\rightarrow\infty$.

The commutator of $p_i$ and $q^j$ looks as follows,
\begin{multline}\label{commut}
  [p_i,q^j]\ket{n}=
  -{\ii\hbar}\delta_i^j\times\\
  \frac{N}{4\pi}\left(\left\{\sin\frac{2\pi n^j}{N_j}-\sin
  \frac{2\pi(n^j+1)}{N_j}\right\}\ket{n+e_i}\right.\\
  \left.+\left\{\sin\frac{2\pi(n^j-1)}{N_j}\right\}\ket{n-e_i}\right)\\ \equiv
  -\ii\delta_i^j\hbar_j\I_{(j)}(N)\ket{n},
\end{multline}
where we have introduced the notation $\I_{(j)}(N)$ for the
following matrix,
\begin{multline}\label{I}
  \I_{(j)}(N)\ket{n}=-\frac{N_j}{4\pi}\left(\left\{\sin\frac{2\pi n^j}{N_j}-\sin
  \frac{2\pi(n^j+1)}{N_j}\right\}\ket{n+e_i}\right.\\
  +\left.\left\{\sin\frac{2\pi(n^j-1)}{N_j}\right\}\ket{n-e_i}\right).
\end{multline}

As it is not difficult to see the equations of motion are not
satisfied by such background. It still can be shown that the
$N=\infty$ solution (\ref{eq:B}) \emph{can not} be approximated by
\emph{solutions} to equations of motion at a \emph{finite} value
of $N$, because at finite $N$ the only solutions are those with
zero commutator, $[A_\mu,A_\nu]=0$.

For (sequences of) vectors $\ket{f}=\sum_nf_n\ket{n}$, on which
$p$ and $q$ remain bounded as $N$ goes to infinity,
\begin{equation}\label{bounded}
  \bra{f}(p^2+q^2)\ket{f}\leq C,
\end{equation}
where $C$ does not depend neither on $N$ nor on $L$ (or $a$),
operator $\I_{(i)}(N)$ approaches the unity one,
\begin{multline}\label{I(N)->I}
  \I_{(j)}(N)\ket{f}\approx \\ \sum_{n}
  \frac{1}{2}\left(f_{n+e_i}\cos\frac{2\pi(n^j-1)}{N_j}+
  f_{n-e_i}\cos\frac{2\pi(n^j+1)}{N_j}\right)\ket{n}\rightarrow
  \ket{f},
\end{multline}
since in this case $n^j\ll N_j$, and $f_{n\pm e_i}-f_n=O(N^{-1})$.

The last equation means that the operators preserving the property
(\ref{bounded}) tend to commute with $\I_{(j)}(N)$ as $N$
approaches the infinity.

\section{Fermionic Contribution}
We are ready now to proceed to the analysis of the fermionic
contribution to the partition function of the model with action
(\ref{action}). Integration over fermionic matrices results in the
Pfaffian of the fermionic operator,
\begin{equation}\label{Pf}
  Z(A)=\int \d\psi \e^{-\Sf},
\end{equation}
where, $\Sf$ stands for the fermionic part of the action
(\ref{action}).

For finite $N$ consider the bosonic background given by matrices
$p_i, q^i$ from eqs. (\ref{qi},\ref{pn}). An arbitrary Hermitian
matrix fluctuation around the given background is
$A_\mu=p_\mu+ga_\mu$. The fermionic part $\Sf$ of the action in
this case looks as follows,
\begin{equation}\label{sf}
  \Sf=-\frac{1}{2}\tr \bar{\psi}\Gamma^\mu [(p_\mu+ga_\mu),\psi],
\end{equation}
where we rescaled $\psi\rightarrow g\psi$.

The free ($a_\mu=0$) part of the fermionic action can be written
in the representation of $\ket{n}$. It looks as follows,
\begin{multline}\label{Sfn}
  \Sf=\sum_{m,n,i} \left((-{\ii\hbar_i}/(2a))\bar{\psi}_{n,m}\Gamma^i
  (\psi_{m+e_i,n}-\psi_{m-e_i,n}-\psi_{m,n+e_i}+\psi_{m,n-e_i})\right. \\
  \left.+L_i\bar{\psi}_{n,m}\bar{\Gamma}_i\left(\sin \frac{2\pi m^i}{N_i}-
  \sin \frac{2\pi n^i}{N_i}\right)\psi_{m,n}\right),
\end{multline}
where,
\begin{align}\label{psi}
  &\psi_{n,m}=\bra{n}\psi\ket{m}, \\
  &\bar{\psi}_{m,n}=\bra{m}\bar{\psi}\ket{n},
\end{align}
and $\Gamma^i$, $\bar{\Gamma}_i$ are  Dirac matrices,
\begin{equation}\label{Dirac}
  \Gamma^ip_i+\bar{\Gamma}_iq^i\equiv \Gamma^\mu p_\mu.
\end{equation}

Although, the action (\ref{Sfn}) differs from the naive lattice
fermionic action, it shares many common features with it.

As for naive lattice fermions, they are described by the action
\cite{Wilson:1974sk},
\begin{equation}\label{naive}
  \Sn=\frac{\ii}{2}\sum_n \bar{\psi}_n
  \Gamma^\mu(\psi_{n+e_i}-\psi_{n-e_i}),
\end{equation}
where for shortening notations we put lattice spacing $a$ to
unity.

A well-known fact is that the actual model suffers from the
fermionic spectrum doubling. The last manifests in the fact that
for each chiral fermionic state there is always another one
present in the spectrum with the opposite chirality but with other
quantum numbers coinciding with the original state. This
phenomenon can be described by introducing some discrete symmetry
which relates these states. The states obtained by action of this
symmetry are called doublers. It is clear that if such a symmetry
exists it completely destroys the chiral asymmetry.\footnote{The
absence of an explicit symmetry of such kind, however, does not
prove necessarily, the absence of doubling.}

In the case of action (\ref{naive}), such a symmetry indeed exists
and its generators in a even dimension $D$ look as follows
\cite{Makeenko:1996bk},
\begin{equation}\label{talph}
  T_\alpha=\ii \Gamma_{(D+1)}\Gamma_\alpha (-1)^{n_\alpha},
\end{equation}
where $\Gamma_{(D+1)}$ is the $D$-dimensional analog of the Dirac
$\gamma_5$-matrix ($D$ is even), $\Gamma_{(D+1)}=\epsilon_D
\Gamma^1\Gamma^2\dots\Gamma^D$, $\epsilon$ is chosen to be either
$\ii$ or $1$ in order to make $\Gamma_{(D+1)}$ Hermitian.

Finding the order of the discrete group generated by $T_\alpha$,
one finds that the number of doubling states is $2^D-1$.

Now, let us return back to to the matrix model given by eq.
(\ref{Sfn}) and try to find a similar symmetry in this case.

One can check that the action (\ref{Sfn}) is left invariant by the
following symmetry,
\begin{equation}\label{doub}
  \psi_{m,n}\rightarrow \ii
  \Gamma_{11}\Gamma_i(-1)^{n^i-m^i}\psi_{m,n}.
\end{equation}
or in the matrix form,
 $$
   \psi\rightarrow T_i\cdot\psi=
   \ii \Gamma_{11}\Gamma_iU_i^{-1}\psi U_i, \eqno
   (\ref{doub}')
 $$
where unitary matrix $U_i$ is given by
$U_i=(-1)^{\frac{N_i}{2\pi}\arcsin (q^i/L_i)}$.

Indeed, factor $\Gamma_{11}\Gamma_i$ commute with all
$\bar{\Gamma}_i$ and $\Gamma^j$ for $j\neq i$ while factors
$(-1)^{n^i-m^i}$ are the same for both $\psi$ and $\bar{\psi}$,
and, therefore cancel. In the remaining term $\Gamma_{11}\Gamma_i$
anticommutes with $\Gamma_i$, but the extra minus sign is
compensated by the variation of the factor $(-1)^{n^i-m^i}$. Thus,
all terms in the action (\ref{Sfn}) remains invariant under the
transformation (\ref{doub}).

Interchange $p\leftrightarrow q$ gives the remaining symmetries,
\begin{equation}\label{doub2}
  \psi \rightarrow \bar{T}_i\cdot\psi=
  \ii \Gamma_{11}\bar{\Gamma}_i\bar{U}_i^{-1}\psi \bar{U}_i,
\end{equation}
where $\bar{U}_i$ is the unitary transformation,
$\bar{U}_i=(-1)^{\frac{N_i}{2\pi}\arcsin (pa_i/\hbar_i)}$.

Summarising one has the action (\ref{action}) invariant with
respect to discrete symmetry generated by $T_\mu$,
\begin{equation}\label{tmu}
  T_\mu\cdot\psi=\ii \Gamma_{11}{\Gamma}_\mu{U}_\mu^{-1}\psi
  {U}_\mu,
\end{equation}
where $U_\mu$ satisfy,
\begin{equation}\label{umu}
  U_\mu^{-1}p_\nu U_\mu=(1-2\delta_{\mu\nu})p_\nu.
\end{equation}

As in the case with naive lattice fermions these transformations
act in such a way that in the continuum limit the states become
$2^D$-fold degenerate with half of that for each chirality. In
particular eq. (\ref{doub2}) means that for each matrix state of
given chirality which connects $m$ and $n$ there are $2^{D/2}$
states of different chiralities connecting $\frac{N_i}{2}e_i-m$
with $\frac{N_i}{2}e_i-n$, $\frac{N_i}{2}e_i+\frac{N_j}{2}e_j+m$
with $\frac{N_i}{2}e_i+\frac{N_j}{2}e_j+n$, $i\neq j$, and so on.
Eq. (\ref{doub}) have the same interpretation in the ``momentum
space'' spanned by the eigenvectors of $p$.

It is clear now that if one wants to compute the gauge anomaly,
one will have contributions from doublers of both chiralities
which cancel each other.

So far, we considered the interaction free part of the fermionic
action. Presence of interaction at least in the framework of
perturbation theory does not change the situation, in this case
one has in the continuum limit an interacting $2^D$-plet instead
of a free one. Let us note, that this analysis may not remain true
beyond the perturbation theory, as for strong field $a_\mu$ the
interaction part dominates and the symmetry (\ref{tmu}) of the
free part does not play in this case such an important role.
Unlike the usual lattice models the study of doubling in
nonperturbative regime is too complicate, but the perturbative
considerations are enough to doubt the result of naive continuum
limit.

One may ask, what happens in this case to the supersymmetry of the
original matrix model given by the action (\ref{action})?

In fact, in spite of the fermionic doubling, the supersymmetry
still exists as it is valid for any matrix configuration and is
irrelevant to the chosen representation. The apparent paradox with
doubled number of fermions is solved if one see that the number of
bosonic ``degrees of freedom'' is also doubled. The bosonic
doublers are gauge equivalent configurations and are related by
gauge transformations $a_\mu\rightarrow U_\nu^{-1}a_\mu U_\nu$.

\section{Doubling States Removing}
The results of the previous section show that matrix model
(\ref{action}) fails to reproduce a chiral continuum model in the
limit $N\rightarrow\infty$. Thus, in particular, one can not
obtain from it the noncommutative SYM model in this limit.

Such a situation can be interpreted either as a presence of finite
$N$ artefact which does not decouple in the limit
$N\rightarrow\infty$, and must be removed by additional effort
like in traditional lattice models, or as an indication that the
model possesses nontrivial symmetries. In traditional lattice
models the doubling states should be removed since the doubling
contradict the continuum ``phenomenology'', in special the
chirality properties of the model.

In the case of matrix models which pretend to describe M-theory
the situation is different. As it is known the M-theory unites
perturbative models with different field content. In particular
IIA models contain states with both chiralities while in IIB
models there are only ones with definite chirality. The duality
symmetry which must relate these models should contain a mechanism
which flips the chirality of states. As we see, such a mechanism
exists on compact noncommutative spaces and is given by doubling.

In spite of this perspective for doubling states to describe the
physical reality there exists, however, possibility to remove them
in order to get a chiral model in the continuum limit. Consider
briefly the ways one can do this. As the problem is a ``lattice''
one, we can look for specific lattice solutions. In the lattice
case the doubling is cured by addition of a Wilson term to the
naive lattice action \cite{Wilson:1974sk}. The problem is that
there is no gauge invariant Wilson term which could be added to
our ``naive'' fermionic action (\ref{sf}), as there is no gauge
and Lorentz invariant fermionic mass term in the model.

One can, however, write down Wilson terms preserving either of two
symmetries.

A possible gauge non-invariant Wilsonian prescription is given by
addition to the naive action (\ref{sf}) of the following term,
\begin{equation}\label{wils1}
  \Delta S_{\mathrm{W,gauge}}=-\frac{1}{2}\tr \bar{\eta}
  \Gamma^\mu[p_\mu,\eta]+[p_\mu^{(+)},\bar{\eta}][p_\mu^{(-)}\psi]+
  [p_\mu^{(+)},\bar{\psi}][p_\mu^{(-)}\eta],
\end{equation}
where $\eta$ is $U(N)$-singlet Majorana-Weyl spinor matrix, and
$p_\mu^{(\pm)}$ are respectively forward and backward one-step
scaled lattice derivatives,
\begin{align}\label{p+-}
  &p_i^{(\pm)}\ket{n}=\pm\ii\sqrt{\frac{\hbar_i}{a_i}}
  (\ket{n\pm e_i}-\ket{n}) \\
  &q^{(\pm)i}\ket{n}=\ii \sqrt{L_i}\e^{\pm 2\pi\ii n^i/N_i}\ket{n}.
\end{align}
Due to the term (\ref{wils1}) the states with large phase of $q$
and $p$, (i.e. with $n^i\sim N_i$ and
$k_i\sim\frac{\hbar_i}{a_i}$) acquire large masses and decouple in
the limit $N\rightarrow\infty$, as it happens in the case with
usual Wilson term.

Another possibility is given by that in contrast to low-dimensional field theory
models where the Lorentz invariance is obligatory, in the Matrix model it is less
important. Moreover, its breaking to lower dimensional symmetries is desirable if
one wants to describe a four dimensional theory in low energy limit, as it was
proposed in the Ref.~\cite{NV}. It is not difficult to construct Wilson term which
breaks, say Lorentz group SO(10) down to SO(9), but preserves the gauge symmetry.
It looks as follows,
\begin{equation}\label{wilson2}
  \Delta S_{\mathrm{W,Lorentz}}=-\frac{1}{2}\tr
  [p_\mu^{(+)},\bar{\psi}]\Gamma^9[p_\mu^{(-)},\psi],
\end{equation}
where $\Gamma^9$ is the 9-th ten dimensional Dirac gamma matrix.

Under this choice modes with large $n$ and $k$ also acquires large
masses in the limit $N\rightarrow\infty$, but it produces terms
which are not invariant with respect to rotations involving the
9-th axis. The gauge symmetry here remains intact. The last should
not appear strange, because giving up a part of Lorentz invariance
in an anomalous model allows one to cancel anomaly. Thus, in the
simplest case of Abelian gauge anomaly in $D=4$ one can cancel the
anomaly $\sim
\epsilon^{\mu\nu\lambda\sigma}F_{\mu\nu}F_{\lambda\sigma}$, where
$\mu,\nu,\dots=0,1,2,3$ by addition a local counterterm $\sim
A_0\epsilon^{ijk}A_i\pd_jA_k$, with $i,j,\dots=1,2,3$ which is not
invariant with respect to Lorentz boosts.

\section{Discussions}
We have shown that perturbative fluctuations in IIB matrix model
at finite $N$ exhibit a phenomenon similar to one in lattice gauge
theories with fermions, consisting in doubling of the fermionic
spectrum.

We considered a simple example of a background given by lattice shift/mo\-men\-tum
shift operators. Basing on analogy with lattice model, we conjecture that this is
a universal feature appearing for arbitrary choice of background configuration
$p_\mu$ which is Hermitian, nondegenerate, etc., and can not be eliminated without
breaking either Lorentz or gauge invariance. The actual results are
straightforwardly translated to the case of the genuine finite $N$ vacua
considered in \cite{Sochichiu:2000eu}.

We give prescriptions for the elimination of the doubling states
in the limit $N\rightarrow\infty$, but preserving either Lorentz
or gauge invariance and, respectively, breaking another one. This
prescriptions should break the supersymmetry, since they do not
restrict the bosonic spectrum as well.

However, in contrast to lattice gauge models, one may \emph{not
need} to eliminate such doubling states. Since there is a
conjecture that IKKT matrix model nonperturbatively describes both
IIB and IIA string models \cite{ikkt}, which have different
chirality content, it may be possible that the doubling in the
language of matrix models is related to the string duality.
\begin{ack} 
I am grateful to the members of our informal seminar for useful
discussions and comments.
\end{ack}

\begin{thebibliography}{10}

\bibitem{ikkt}
N.~Ishibashi, H.~Kawai, Y.~Kitazawa, and A.~Tsuchiya, ``A
large--{N} reduced model as superstring,'' {\em Nucl. Phys.} {\bf
B498} (1997) 467, {{\tt hep-th/9612115}}.

\bibitem{Banks:1997vh}
T.~Banks, W.~Fischler, S.~H. Shenker, and L.~Susskind, ``M theory
as a matrix model: A conjecture,'' {\em Phys. Rev.} {\bf D55}
(1997) 5112--5128, {{\tt hep-th/9610043}}.

\bibitem{Schild:1977vq}
A.~Schild, ``Classical null strings,'' {\em Phys. Rev.} {\bf D16}
(1977) 1722.

\bibitem{Connes:1998cr}
A.~Connes, M.~R. Douglas, and A.~Schwarz, ``Noncommutative
geometry and matrix theory: Compactification on tori,'' {\em JHEP}
{\bf 02} (1998) 003, {{\tt hep-th/9711162}}.

\bibitem{Sochichiu:2000eu}
C.~Sochichiu, ``M[any] vacua of {IIB},'' {\em JHEP} {\bf 05}
(2000) 026, {{\tt hep-th/0004062}}.

\bibitem{Gracia:2000eu}
J.~M. Gracia-Bond\'{\i}a and C.~P. Mart\'{\i}n, ``Chiral gauge
anomalies on {noncommutative} $\mathbb{R}^4$,'' {{\tt
hep-th/0002171}}.

\bibitem{Bonora:2000eu}
L.~Bonora, M.~Schnabl, and A.~Tomasielo, ``A note on consistent
anomalies in noncommutative {YM} theories,'' {{\tt
hep-th/0002210}}.

\bibitem{Creutz:1996wq}
M.~Creutz and M.~Tytgat, ``Species doubling and chiral
lagrangians,'' {\em
  Phys. Rev. Lett.} {\bf 76} (1996) 4671--4674,
{{\tt hep-ph/9605322}}.

\bibitem{Kits}
N.~Kitsunezaki, J.~Nishimura, ``Unitary IIB Matrix Model and the Dynamical
Generation of the Space-Time," {\em Nucl.Phys.} {\bf B526} (1998) 351, {\tt
hep-th/9707162}.

\bibitem{Wilson:1974sk}
K.~G. Wilson, ``Confinement of quarks,'' {\em Phys. Rev.} {\bf
D10} (1974) 2445--2459.

\bibitem{Nielsen:1981xu}
H.~B. Nielsen and M.~Ninomiya, ``Absence of neutrinos on a
lattice. 2. Intuitive topological proof,'' {\em Nucl. Phys.} {\bf
B193} (1981) 173.

\bibitem{Nielsen:1981rz}
H.~B. Nielsen and M.~Ninomiya, ``Absence of neutrinos on a
lattice. 1. Proof by homotopy theory,'' {\em Nucl. Phys.} {\bf
B185} (1981) 20.

\bibitem{NV}
J.~Nishimura, G.~Vernizzi, ``Spontaneous Breakdown of Lorentz Invariance in IIB
Matrix Model,'' {\em JHEP.} {\bf 04} (2000) 015, {\tt hep-th/0003223}

\bibitem{Makeenko:1996bk}
Y.~M. Makeenko, ``Introduction to gauge theory,'' {\em Surveys
High Energ. Phys.} {\bf 10} (1997) 1.
\end{thebibliography}

\end{document}